\documentclass{iau} 
\usepackage{graphicx}

\title[Dwarf Galaxies: Their Low Metallicity Interstellar Medium  ] 
{Dwarf Galaxies: Their Low Metallicity Interstellar Medium}

\author[Suzanne C. Madden and Diane Cormier ]   
{Suzanne C. Madden$^1$ and Diane Cormier$^1$}

\affiliation{$^1$ Departement d'Astrophysique, CEA, Saclay, \\ Gif-sur-Yvette 91400, France \\ email: {suzanne.madden@cea.fr}  diane.cormier@cea.fr \\[\affilskip]}

\pubyear{2018}
\volume{344}  
\jname{Dwarf Galaxies: From the Deep Universe to the Present}
\editors{S. Stierwalt Editor \& K. McQuinn Editor, eds.}
\begin{document}

\maketitle

\begin{abstract}
This review describes where we are today in light of the dust and gas properties and their relation to star formation, in low metallicity galaxies of the local universe following recent surveys from sensitive infrared space telescopes, mainly \textit{Spitzer} and \textit{Herschel} space observatories as well as ground-based observations of the molecular gas reservoir. Models to interpret the ISM properties are gaining sophistication in order to account for the wide range of valuable observational diagnostics that we have today to trace the different gas phases, the broad range of photometry we have, from mid-infrared to submillimetre dust emission and the various galactic size scales that we can sample today. This review summarizes the rich multi-phase observations we can exploit today, and the multi-phase modeling approach to interpret the observations.

\end{abstract}

\firstsection 
\section{Introduction}

A few decades ago, a review on the gas and dust in dwarf galaxies would have been quite brief. Optical observations were successfully revealing the star formation characteristics in
dwarf galaxies, and there was relatively little attention given to the substance of the interstellar material and how it might interact with and shape star formation. Our picture of the interstellar medium (ISM) of low-metallicity dwarf galaxies has changed dramatically since the advent of infrared (IR) space telescopes and more sensitive and larger ground-based telescopes. Metal enrichment and dust and gas accumulation, the processes necessary to feed star formation, are the heartbeat of galaxies from their birth to their evolution throughout cosmic time.  As our knowledge of the early steps of galactic birth and life is making progress and observations of the cold gas and dust are being pushed to the frontiers of cosmic history at an accelerated pace, it is time to validate models we can construct from the wide range of low metallicity environments in the local universe and to eventually test their application to early universe conditions, now that dust and gas are being uncovered in systems as young as 1\,Gyr old (e.g., \cite[Carilli \& Walter 2013]{carilli13}, \cite[Venemans \etal\ 2017]{venemans17}, \cite[Laporte \etal\ 2018]{laporte17}). 
 
Low metallicity dwarf galaxies differ from their more metal-rich counterparts in many ways, being lower in mass, elemental abundance and luminosity, for example. These basic properties can conceivably alter the distribution of the different galactic phases, the propagation of starlight, the physics and chemistry of the gas and dust, the way stars form out of molecular clouds as well as the mechanics of large scale dynamical and gravitational effects. Thanks to more sensitive telescopes over the decades, especially in the IR, we can delve into the workings of these curious systems.
 
 The attention to the ISM properties of dwarf galaxies is widespread by now, due to remarkable technological achievements in telescopes, especially IR airborne and space telescopes and larger and more sensitive ground-based telescopes. Most recently, \textit{Herschel} has brought dwarf galaxies to the stage, with vast wavelength capability to study the energy budget and the structure of the different phases in dwarf galaxies. This review presents where we are today in assessing the properties of the gas and dust in star-forming dwarf galaxies in our local universe, as we look forward to preparing future strategies that will follow up on expanding our vision, both for the local universe dwarf galaxies as well as pushing to deeper cosmological epochs to uncover many early-universe galaxies which may be reaching primordial conditions.

This review is organised in the following way. Section \ref{dust} brings to light the dust observations and their properties primarily from \textit{Spitzer} and \textit{Herschel} observations, and the modeling results, comparing the low metallicity galaxies to their more metal-rich counterparts. We exploit the gas and dust observations for multi-phase modeling on galaxy-wide scales as well as a look into some smaller scale model studies (Section \ref{modeling}). Section \ref{mol} describes where we are in the hunt for evidence of molecular gas and the enigma this brings in confronting the star formation properties of dwarf galaxies with their elusive molecular gas reservoirs.  We then show how it is possible to quantify the molecular gas reservoir via the far infrared (FIR) [CII]158$\mu$m observations. We conclude with a look to the future (Section \ref{conclusion}).

\subsection{Challenges related to observations of dwarf galaxies}
Recent observations of the ISM of star-forming dwarf galaxies have brought detections of objects with metallicities as low as $\simeq$3\% solar (the lowest to date being AGC 198691; \cite[Hirschauer \etal\ 2016]{hirschauer16}) with very intriguing characteristics. Star-forming dwarf galaxies stand out as puzzling objects for several reasons. Since they are forming stars, one would expect cold molecular gas, the fuel for star formation, to be present. However, CO is found systematically to be faint, and remains a challenge at the lowest metallicities, even as millimeter telescopes have become more and more sensitive. The verdict is still out on how star formation proceeds in these low metallicity galaxies. Are they efficient or non-efficient in turning gas into stars (Section \ref{mol})? Perhaps the question is not \textit{how} are some dwarf galaxies vigorously forming stars with little nor no molecular gas detected but, instead \textit {where} is the molecular gas for star formation in dwarf galaxies?

Our local universe dwarf galaxies are systems that are not necessarily young chronologically - they have undergone some ISM recycling. Since their measured elemental abundances are overall lower than their metal-rich counterparts, we can ask: are the physical properties of the dust they harbor different?  One of the perplexing properties of dust emission that is observed to be more prominent in dwarf galaxies is the excess submm emission, the origin of which is still unknown today. Additionally, the expectation is that small dust particles (such as PAHs) would be present at some level in dwarf galaxies. They are very prominent in the mid-infrared (MIR) spectra of spiral galaxies, for example, but very faint, if present at all, in dwarf galaxies. These are some of the features of dwarf galaxies, as well as the fact that in general, the presence of dust becomes scarce in the lowest metallicity galaxies, more so than expected (Section \ref{dust}).

These attributes of dwarf galaxies hint at the fact that the environment in which dust and molecules are formed and live is different than in the Milky Way or other metal-rich environments. Some of the current open questions that we aim to tackle in this review are: To what extent do ISM physical conditions, composition, structure, cooling and heating balance, change in low-metallicity star-forming galaxies? How can we best characterize those ISM properties and how do they impact star formation? What are the total dust and gas budgets of dwarf galaxies and how can we trace them? \\
 
\section{The dust properties of dwarf galaxies}
\label{dust}

Metal enrichment and dynamical processes in the ISM drive the dust life cycle. Stars are born in molecular clouds and nucleosynthesize the elements that go into the building blocks of dust. Supernovae and AGB stars eject these elements into the ISM, some of which remain in the gas phase and some go into forming dust. Dust is processed in the ISM via different dynamical events such as shock waves and stellar winds, which fragment and erode the dust, releasing some elements back into the ISM. Dust is accumulated in molecular clouds, where it shields the gas from ultra-violet (UV) destruction.  Eventually star formation is ignited in the dense molecular gas and the life cycle repeats. Each of these cycles enriches the ISM further with heavy elements. Dwarf galaxies have probably not gone through many of these cycles, as they are not as chemically rich, as our Galaxy, or other spiral galaxies. In this way the dust mass and the dust-to-gas mass ratio (D/G) closely trace the enrichment of the ISM.  The D/G is the relationship between the amount of metals locked up in dust with those present in the gas phase and is governed by the competition between dust creation versus dust destruction in the ISM. It is a powerful tracer of the evolutionary stage of a galaxy.

 While dust is a very minor fraction of the mass in galaxies, it plays a major role in regulating the thermal balance and makes an important link to the gas excitation.  Dust grains are the major players in the heating of photodissociation regions (PDRs), where the MIR and FIR emission lines act as important gas coolants. Thus, dust links tightly with the heating and cooling of galaxies.  As dust couples closely with the gas, and is a good tracer of gas, the mass of dust is often used to estimate the total gas mass of galaxies. 
   
\begin{figure*}
\centering
\begin{minipage}{.5\textwidth}
  \centering
  \includegraphics[width=2.7in]{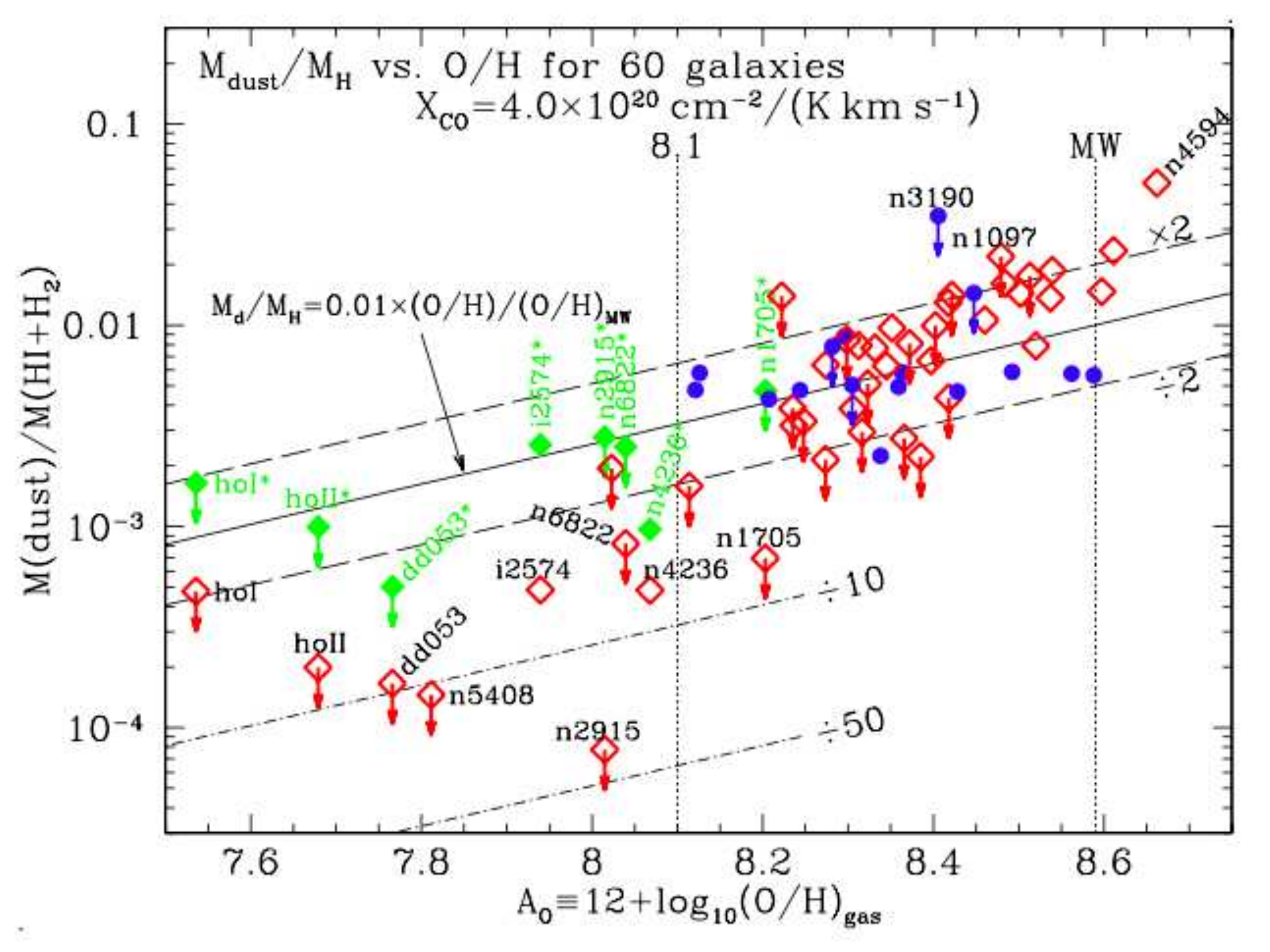}
 \label{fig:test1}
\end{minipage}%
\begin{minipage}{.5\textwidth}
  \centering
    \includegraphics[width=2.0in, angle=-90]{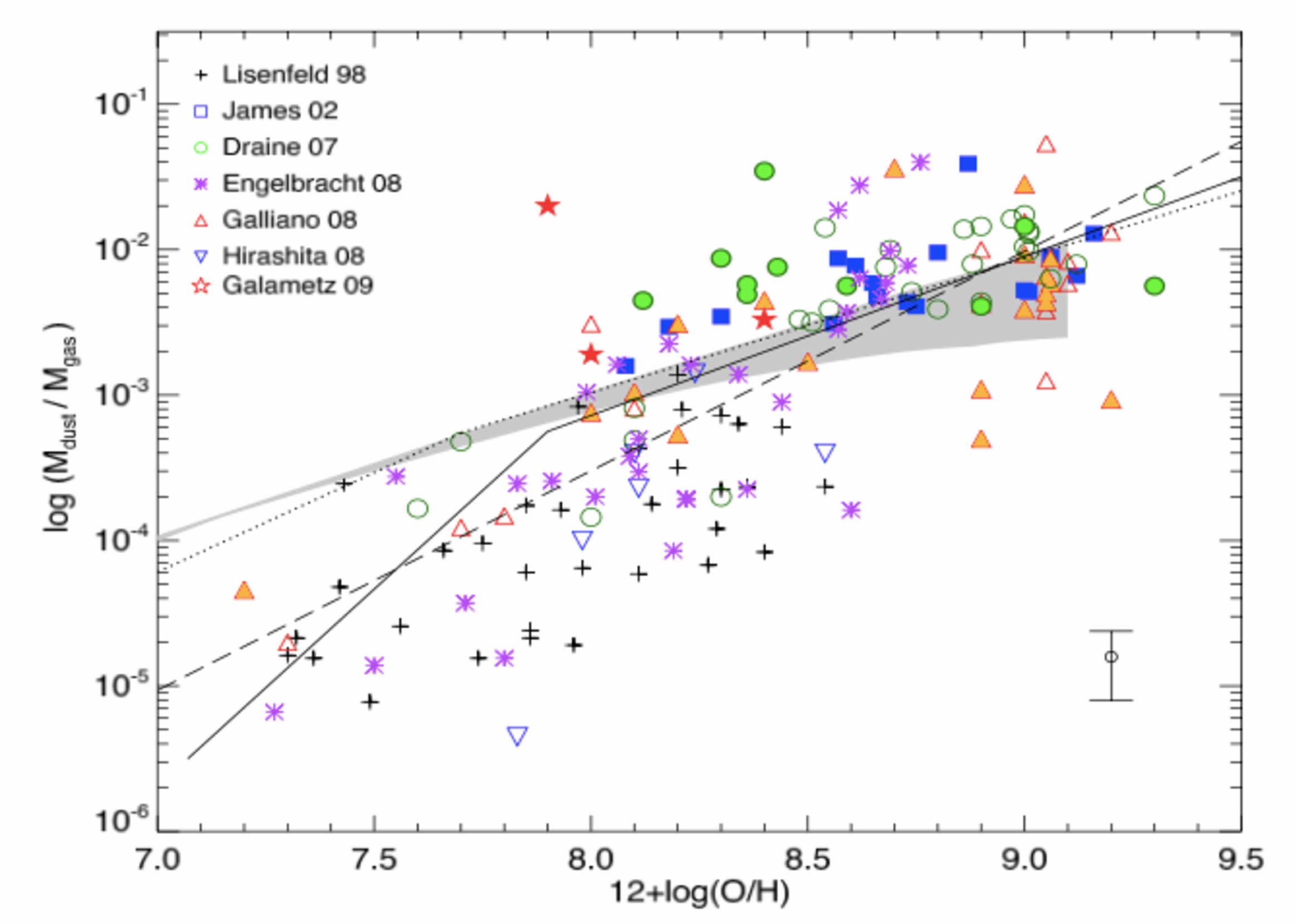}
  \label{fig:test2}
\end{minipage}
\caption{{\it Left:}  D/G measured as a function of metallicity for the SINGS galaxies (\cite[Draine \etal\ 2007]{draine07}). Galaxies with circles  have 850$\mu$m SCUBA data used in the SED modeling. Diamonds designate those galaxies without SCUBA data. Filled vs. open diamonds show D/G when the HI and IR are measured in the same apertures vs. measured IR and total  galactic HI. {\it Right:} D/G as a function of metallicity for numerous surveys. Filled symbols indicate the use of submm data in the SED modeling (\cite[Galametz \etal\ 2011]{galametz11}).  The lines are different models: solid line from \cite{edmunds01}; dotted line from \cite{dwek98}; dashed line is the linear regression of the full sample.  }
\label{DG_MG_BD}
 \end{figure*}
 
\subsection{How do dust properties of dwarf galaxies compare to metal-rich galaxies?}
\label{dust_properties}

It was only with \textit{IRAS}, launched 25 years ago, that the windows into the MIR to FIR wavelengths provided the first glimpse into dust emission properties of dwarf galaxies with pioneering studies (e.g. \cite[Hunter \etal\ 1986]{hunter86}, \cite[Melisse \etal\ 1994]{melisse94}, \cite[Israel \etal\ 1996]{israel96}). It was already becoming apparent, even with the wavelength coverage of \textit{IRAS} limited to 100$\mu$m, that star-forming dwarf galaxies tend to harbor overall warmer dust than spiral galaxies.  Already with the limited data available for dwarf galaxies, it hinted that D/G was dropping more rapidly than expected toward lower metallicity (\cite[Lisenfeld \& Ferrera 1998]{lisenfeld98}, \cite[James \etal\ 2002]{james02}).  \textit{ISO} and \textit{Spitzer} advanced the subject further, with a wider rage of broad bands, as well as spectroscopy, revealing the dearth of MIR PAH bands in low metallicity dwarf galaxies (e.g. \cite[Galliano \etal\ 2003]{galliano03}, \cite[Galliano \etal\ 2005]{galliano05}, \cite[Madden \etal\ 2006]{madden06}, \cite[Engelbracht \etal\ 2008]{engelbracht08}, \cite[Rosenberg \etal\ 2008]{rosenberg08}).  

Extending to 160$\mu$m, \textit{Spitzer} gave us a hint of the presence of cooler dust in dwarf galaxies as well as well-sampled MIR dust emission. The fact that surveys of galaxies could now reach lower metallicities allowed investigations of a wide parameter space of star formation properties, luminosity, metallicity, size scale, extinction properties, etc.  \cite{draine07} modeled the dust properties in the SINGS galaxies observed with 
\textit{Spitzer}, extending to metallicities where FIR/submm detections made it possible to determine D/G  to 12+log(O/H) $\sim$ 7.5, finding that the mass fraction of PAHs is correlated with metallicity. Given the upper limits of the observed D/G of the lower metallicity galaxies, this study suggested that for the lower metallicity galaxies, the D/G may indeed deviate from the linear behavior of D/G with metallicity, like the higher metallicity galaxies follow (Figure \ref{DG_MG_BD}). The study of \cite{galametz11} confirmed a non-linear trend in the D/G at the lowest metallicities (Figure \ref{DG_MG_BD}) and, with complementary ground-based LABOCA and SCUBA submm observations, found a significant increase in dust mass was present in some dwarf galaxies, considering the colder dust being sampled in the submm.

\begin{figure*}
\begin{center}
\hspace*{-2.0in}\includegraphics[width=5.5in]{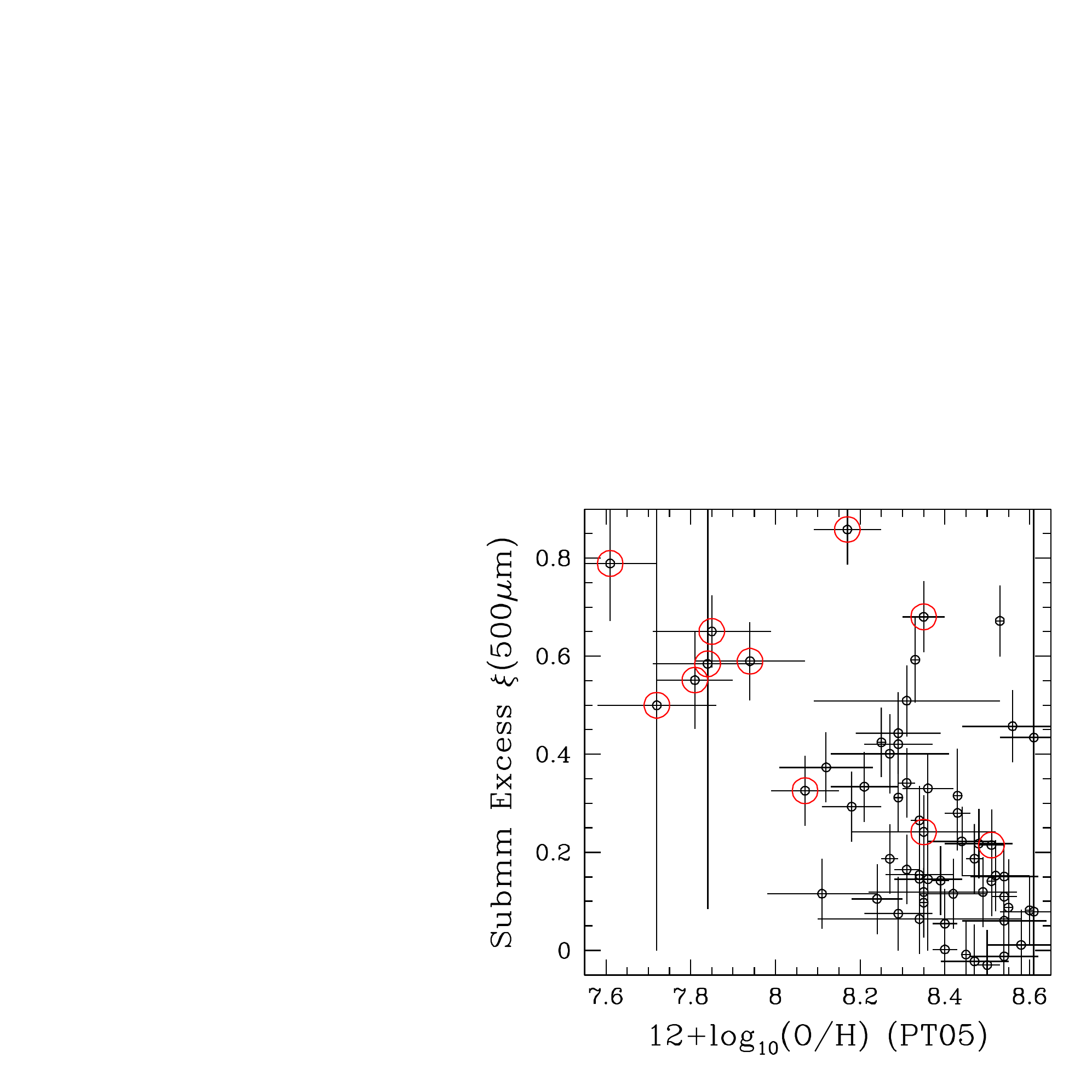} 
%
 \caption{Submm excess, emission above that which can be accounted for in the dust SED model, as a function of metallicity (\cite[Dale \etal\ 2017]{dale17}). The submm excess shows a preference toward  lower metallicity environments. Red circles indicate irregular dwarf galaxies.}
\label{submm_excess}
\end{center}
\end{figure*}

\textit{Herschel} revolutionised the field, extending the wavelength range out to 500$\mu$m, pinning down the observed MIR to submm spectral energy distributions (SEDs), allowing accurate dust masses to be determined.  The \textit{Herschel} Dwarf Galaxy survey (DGS; \cite[Madden \etal\ 2013]{madden13}) mapped 50 dwarf galaxies of the local universe to quantify the effect of the metallicity on the gas and dust properties, spanning a wide range of metallicities, star formation rates, and size scales. \cite{remy15} compared the properties of the low-metallicity DGS galaxies with those of the overall more metal-rich survey, KINGFISH (\cite[Kennicutt \etal\ 2011]{kennicutt11},  \cite[Dale \etal\ 2012]{dale12}). Modeling 109 galaxies with the \cite{galliano11} dust SED model, and determining metallicities self-consistently, comparison revealed notable differences between metal-rich and metal-poor galaxies spanning 2 dex in metallicities: 1) systematically higher dust temperatures ($T_{\rm dust}$) in the low-metallicity galaxies: $T_{\rm dust}$ $\sim$ 30K while the more metal-rich galaxies have colder dust: $T_{\rm dust}$ $\sim$ 20K ; 2) broader FIR peaks in the SEDs, suggesting clumps with broad ranges of average dust temperatures; 3) faint or absent MIR PAH features, thought to suffer destruction in the harsh environment; 3) significantly lower FIR luminosities ($L_{\rm FIR}$), star formation rates (SFR) and dust masses and 4) tendency for submm excess emission, not fitted by the SED model, having a preference for low metallicity galaxies, also found by \cite{galametz11} and shown in \cite{dale17} (Figure \ref{submm_excess}).

\begin{figure*}
\centering
\begin{minipage}{.4\textwidth}
  \centering
  \includegraphics[width=2.2in]{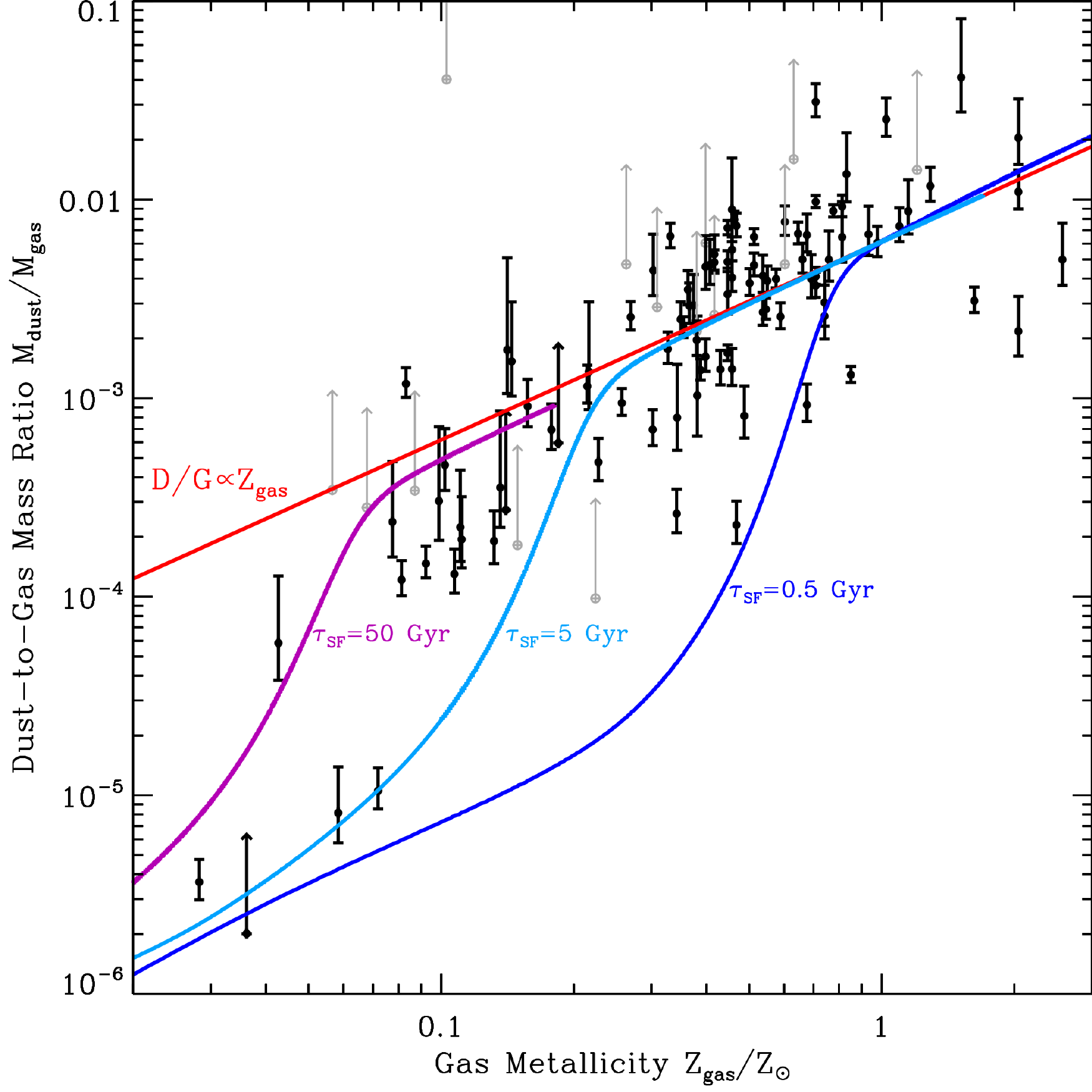}
\end{minipage}%
\begin{minipage}{.6\textwidth}
  \centering
  \includegraphics[width=3.0in]{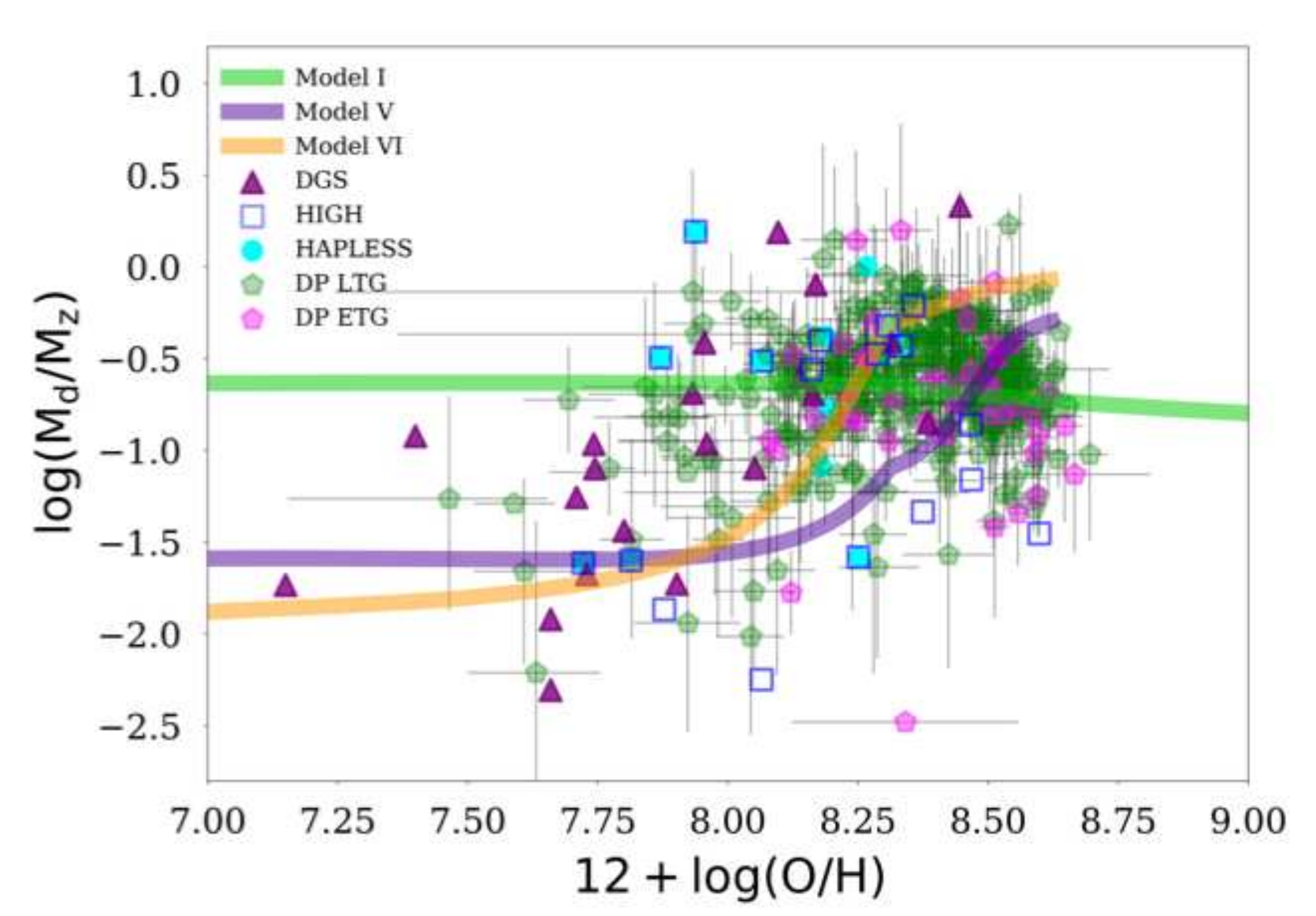}
\end{minipage}
\caption{{\it Left:} Variations of the D/G ratio as a function of metallicity, for galaxies from the Herschel DGS and KINGFISH surveys (\cite[R\'emy-Ruyer \etal\ 2014]{remy14}). The average trend is linear until around $12+\log(O/H) \le 7.8$ (Z $\le $ 0.1 Z$_\odot$). Predictions from chemical evolution models (\cite[Asano \etal\ 2013]{asano13}) are plotted for different star formation time scales, explaining the spread of D/G and the drop at low metallicities. {\it Right:} Dust-to-metals ratio vs. metallicity for a wide variety of surveys. The metals seem to be produced while the dust mass is relatively diminished for the low metallicity galaxies. The low metallicity galaxies come from the Dwarf Galaxy Survey (DGS) and the DustPedia galaxies (DP) (\cite[De Vis \etal\ 2018]{devis18})  Different chemical evolution models are overlaid on the data.}
\label{DG_ARR_PDV}
\end{figure*}

Due to the superior sensitivity of \textit{Herschel}, along with its extensive wavelength windows opening up the submm regime, it is now possible to extend D/G measurements to the lowest metallicities, reaching two extremely low metallicity galaxies in the local universe, SBS0335-052 and IZw18. Figure~\ref{DG_ARR_PDV} shows that the D/G is roughly linear with metallicity until around 10\% solar metallicity (12+log(O/H) $\sim$ 8.0) and then decreases in a nonlinear fashion, becoming much steeper at lower metallicities (\cite[R\'emy-Ruyer \etal\ 2014]{remy14}, \cite[De Vis \etal\ 2017]{devis17}, \cite[De Vis \etal\ 2018]{devis18}). At metallicities as low as a few percent that of solar, dust becomes a  rare component with 2 orders of magnitude lower dust mass than that expected if metals are condensed into dust in a more efficient way, as in the more massive galaxies, such as our Galaxy. Elements are not incorporated into dust in the same manner at lower metallicities.
 
Chemical evolution models of \cite{asano13}, \cite{zhukovska14}, \cite{devis17}, \cite{devis18} attempt to explain the D/G vs. metallicity behavior. Dust growth in the ISM is a necessary dust production process to include in the models. Variation in the star formation time scales can account for both the scatter at a fixed metallicity as well as the drop in D/G at the lower metallicities (Figure~\ref{DG_ARR_PDV}). The sources of dust, such as supernovae, are producing the elements, but the process of accretion of the elements into dust necessitates the presence of a dense phase which requires a longer time to "grow" in the low-metallicity galaxies. Thus galaxies begin with low dust-to-metals (D/Z) and as they evolve, the D/Z increases as dust grains have the right conditions to grow - being balanced at later evolutionary stages by the dust grain growth and dust destruction.
 
This non-linearity in the D/G raises a cautionary note to studies which use an assumed D/G when measuring dust mass, to get to the gas mass. Even for higher metallicity galaxies, there can be at least an order of magnitude uncertainty due to the wide scatter in the D/G as a function of metallicity. Additionally, if the source is low metallicity, the gas mass can be underestimated if the D/G is assumed to be a linear relation with metallicity. In the future, more sensitive telescopes, such as \textit{SPICA}, will hopefully, be able to measure dust masses for a larger number of dwarf galaxies and a wider range of metallicities to get a better handle of the D/G at the lower metallicities and to understand the driving elements in the chemical evolution of galaxies.

For a more detailed review on the dust properties of galaxies, including low metallicity galaxies, see \cite{galliano18}.

\section{Gas properties via multi-phase modeling of the ISM of dwarf galaxies}
\label{modeling}
Along with the cold dust, recent observations from IR and millimeter(mm) observatories have unveiled properties of the gas that resides in more embedded and/or cold phases surrounding star-forming regions (e.g., \cite[Hunt \etal\ 2010]{hunt10}, \cite[Madden \etal\ 2013]{madden13}, \cite[Cigan \etal\ 2016]{cigan16}). The physics and chemistry of those regions is mainly driven by UV photons that can travel and affect the ISM to larger distances in dust-poor environments. Modeling these ISM phases is therefore of great interest to understand how ISM properties (physical conditions, structure, energetics) vary with the metallicity of a galaxy. \\
 
\subsection{What does the ISM look like globally and locally?}
\label{ISM correlations}
\begin{figure}[h]
\begin{center}
 \includegraphics[clip,trim=12mm 10mm 10mm 0,width=2.2in]{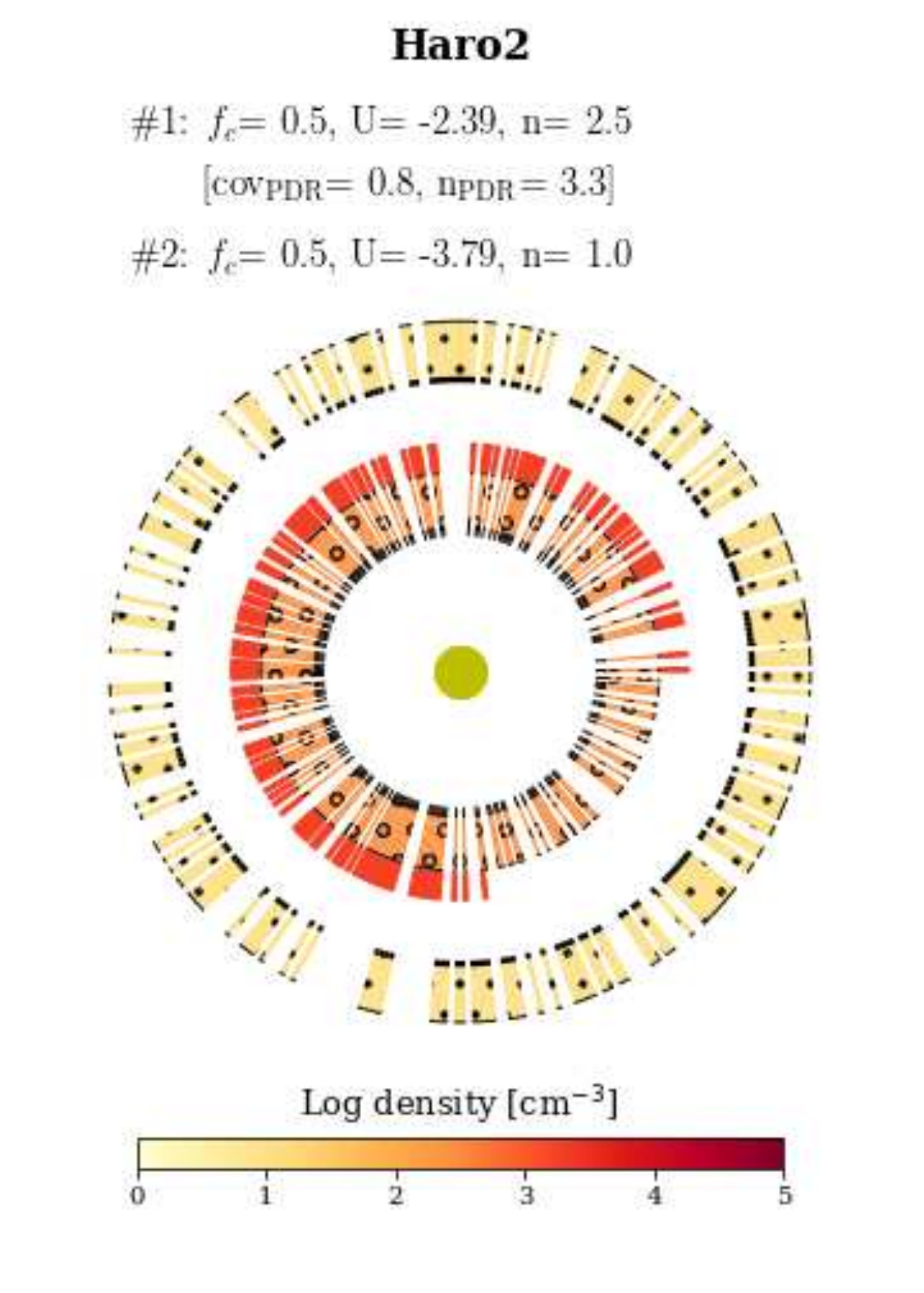} \hfill
 \includegraphics[clip,trim=12mm 0 10mm 0,width=2.6in]{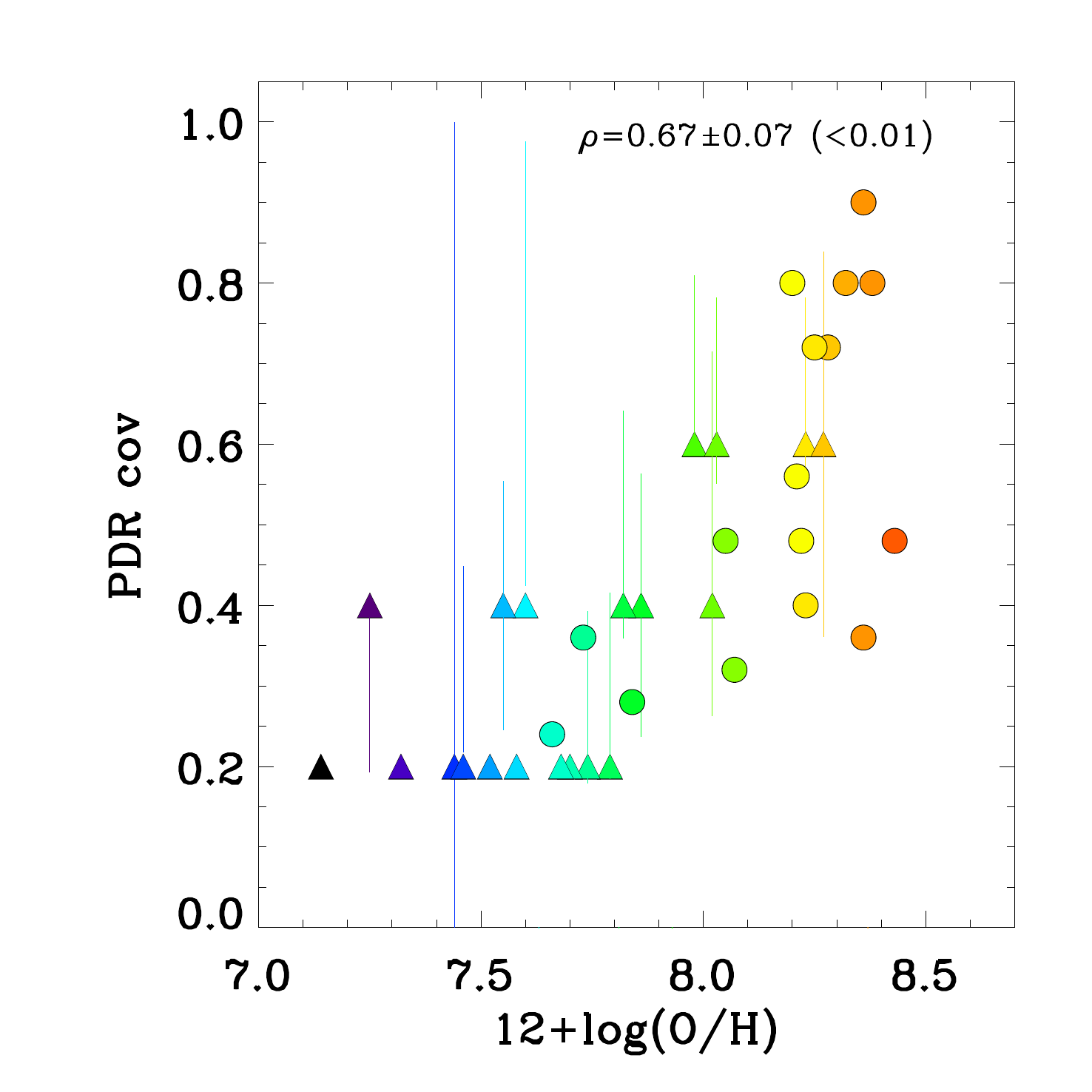}
 \caption{{\it Left:} Schematic of the HII region and PDR model of the dwarf galaxy Haro\,2 using Cloudy. Model with 2 components: f$_c$ is the covering factor of the ionized gas components of density (n) and ionization parameter (U). PDR component (solid annular component): cov$_{PDR}$ is the PDR covering factor and n$_{PDR}$ is the density of the PDR. {\it Right:} Correlation between the covering factor for the PDR and metallicity obtained after modeling each galaxy of the DGS (\cite[Cormier \etal\ 2018]{cormier18}).}
\label{fig:model}
\end{center}
\end{figure}

\
Observations of MIR and FIR spectral lines in nearby star-forming dwarf galaxies, and their modeling with spectral synthesis codes such as Cloudy (\cite[Abel \etal\ 2005]{abel05}, \cite[Ferland \etal\ 2017]{ferland17}), have revealed dramatic differences between the ISM of metal-rich and metal-poor galaxies. We highlight a few of these differences here:
\begin{itemize}
\item
relatively bright ionic lines, especially of species with high ionization potentials (e.g., [SIV]10$\mu$m, [NeIII]15$\mu$m, [OIII]88$\mu$m). Ratios of [SIV]/[SIII], [NeIII]/[NeII], [OIII]/[NII] are higher in metal-poor galaxies by a factor of 10 typically, indicative of intense, hard radiation fields and higher ionization parameters than in normal, metal-rich galaxies (\cite[Hunter \etal\ 2001]{hunter01}, \cite[Madden \etal\ 2006]{madden06}, \cite[Cormier \etal\ 2015]{cormier15}). 
\item
the main PDR coolants, [CII]157$\mu$m and [OI]63$\mu$m, are detected for the first time in a large sample of dwarf galaxies. Their ratios with $L_{\rm TIR}$, on the order of one percent, is sometimes seen as a proxy for the efficiency of heating by photo-electric effect. This heating is rather efficient, which could be due to the dilution of the UV field on large spatial scales (e.g., \cite[Israel \& Maloney 2011]{israel11}, \cite[Cormier \etal\ 2015]{cormier15}). It is also found quite constant on spatial scales of a few parsecs (e.g., \cite[Lebouteiller \etal\ 2012]{lebouteiller12}).
\item
the [OIII]88$\mu$m line is often found to be even brighter than the [CII]157$\mu$m line, unlike in metal-rich galaxies.
Modeling of ionized and neutral gas on whole-galaxy scales indicates that the ionized gas fills a larger volume than in metal-rich galaxies and that the neutral gas has a low covering factor, highlighting an increase of the ISM porosity at low metallicity. Figure~\ref{fig:model} shows a schematic of current models of the ISM (HII region and PDRs) obtained from modeling the line emission of $\sim$20 MIR and FIR lines (\cite[Cormier \etal\ 2012]{cormier12}, \cite[Cormier \etal\ 2018]{cormier18}). By modeling each galaxy of the DGS, we find that the covering factor of the PDR gas correlates with the metallicity of a galaxy (Fig.~\ref{fig:model}). How this covering-factor may be related to escape fraction of ionizing photons, is not yet clear.
\end{itemize}

The UV photons from stellar clusters are believed to be the main energy source in star-forming dwarf galaxies. Evidence for emission from shock tracers (e.g., [FeII] or H$_2$ lines in the MIR range) is weak but requires further, more sensitive observations (JWST). Weak AGN or soft X-rays can also be important sources of heating in individual cases, such as I\,Zw\,18 where dust is not abundant enough to significantly heat the PDR (\cite[Lebouteiller \etal\ 2017]{lebouteiller17}).

\subsection{Modeling the small spatial scales - Local Group galaxies}
\label{spatial models}
Local Group galaxies offer the opportunity to study the ISM on scales of a few parsecs to the size of star-forming complexes, although at the expense of incomplete mapping of galaxies.

The wealth of ancillary and good signal-to-noise data for local galaxies is an asset to constrain the various parameters of the models (heating source, radiation field, density, abundances, equation of state, etc.). One big advantage compared to global studies is that individual heating sources (such as stars) can be identified and their effect on the surrounding gas can be better understood. 
For example, \cite{lee16} analyze the PDR emission and molecular gas excitation via multiple CO lines (as high as J=12-11) and the FIR PDR tracers, [OI], [CII] and [CI] at $\sim$10\,pc scales in the region N\,159 of the LMC (d$\simeq$50\,kpc, Z$\simeq$0.5 solar) to identify which heating source (cosmic rays, X-rays, UV photons, shocks) is dominant. While PDR models can account for the FIR cooling lines, they produce weak CO emission. Ruling out X-rays and cosmic-rays, \cite{lee16} find that low velocity C-type shocks are required to reproduce the warm CO that is observed.
We also refer the reader to the study of N\,11-B in the LMC by \cite{lebouteiller12}.

Moreover, with spatially-resolved observations, the projected ISM distribution is better known, allowing, in principle, for more accurate model representations. 
In this direction, an interesting result stems from the study of the PDR properties in the well-known star-forming region 30\,Doradus in the LMC by \cite{chevance16}. Their PDR modeling on $~$3\,pc scales  has allowed the reconstruction of the 3D geometry of the region by comparing the radiation field intensity from the models to that from the stellar cluster R\,136.

\begin{figure}
\begin{center}
\includegraphics[width=\textwidth]{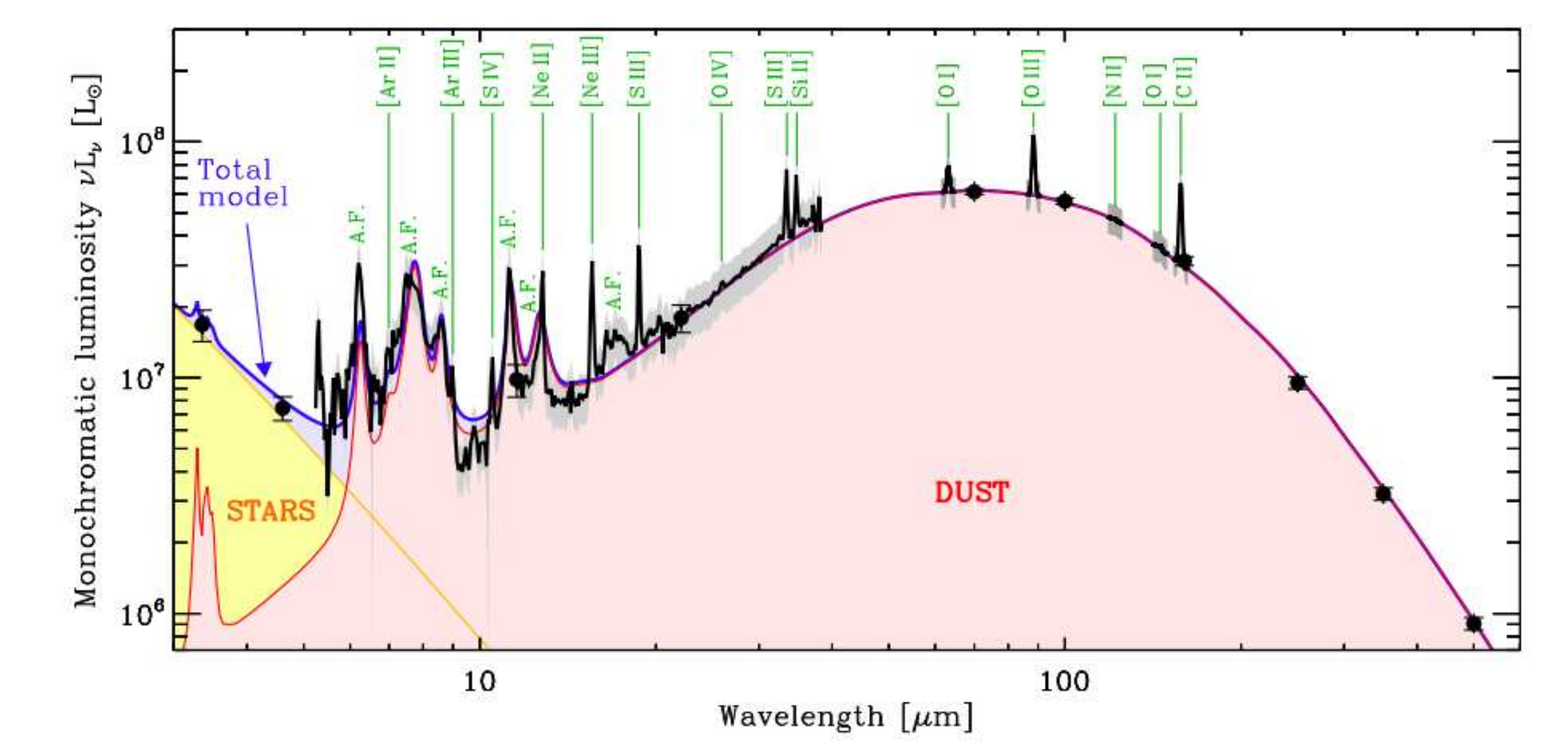}
\hfill 
\caption{The modeled SED of a region of 200 pc x 250 pc, toward the central star-forming region in IC 10 (\cite[Polles \etal\ 2018]{polles18}) using the observations from \textit{WISE}, \textit{Herschel}/PACS and \textit{Herschel}/SPIRE photometry (dots). Included in the SED are the emission lines from the \textit{Spitzer}/IRS and \textit{Herschel}/PACS spectrometers. The continuum covers the total dust component (full wavelength range) and the stellar continuum ($\le$ 10$\mu$m).The bright MIR features include the aromatic features (A.F.) also called PAHs. All of these multi-wavelength observations probe a variety of dust and gas phases.}
\label{ic10}
\end{center}
\end{figure}

Going to larger distances, one can study more broadly the propagation of radiation into the ISM. \cite{polles18} analyze the properties of the ionized gas from \textit{Spitzer} and \textit{Herschel} spectroscopy (Figure \ref{ic10}) in the dwarf irregular galaxy IC\,10 (d$\simeq$700\,kpc, Z$\simeq$0.3 solar). Their modeling is carried out on carefully-selected regions that cover different spatial scales (from 20\,pc to 200\,pc) to investigate how conditions change with scale and discuss possible biases resulting from the modeling method. Interestingly, they show that clumps on small scales are matter-bounded, i.e. letting ionizing photons escape, while on larger scales, the regions become more radiation-bounded.

\section{Molecular gas reservoir}
\label{mol}
 
\begin{figure*}
\centering
\begin{minipage}{.5\textwidth}
  \centering
    \includegraphics[width=2.7in]{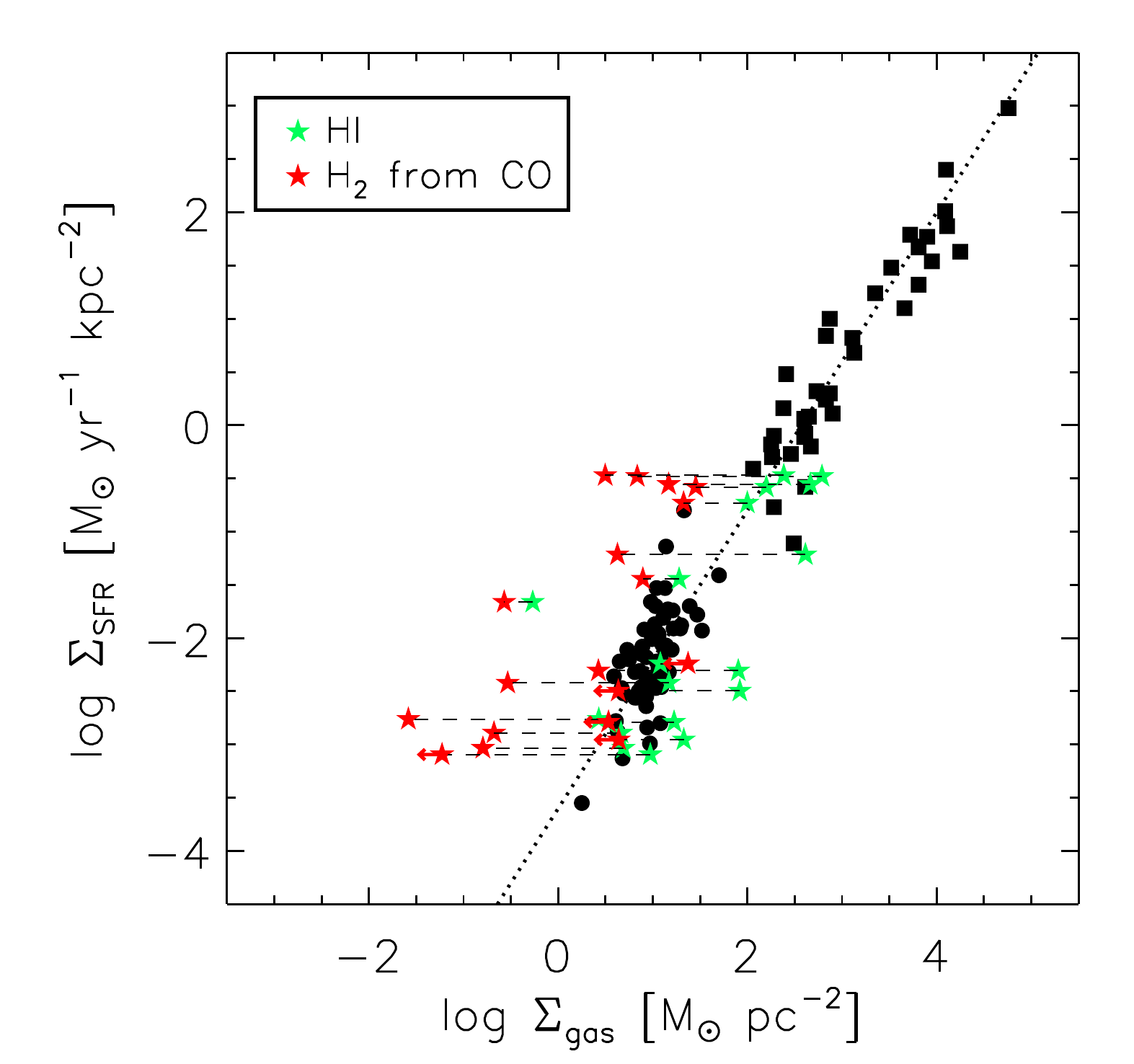}
  \label{fig:test1}
\end{minipage}%
\begin{minipage}{.5\textwidth}
  \centering
  \includegraphics[width=2.8in]{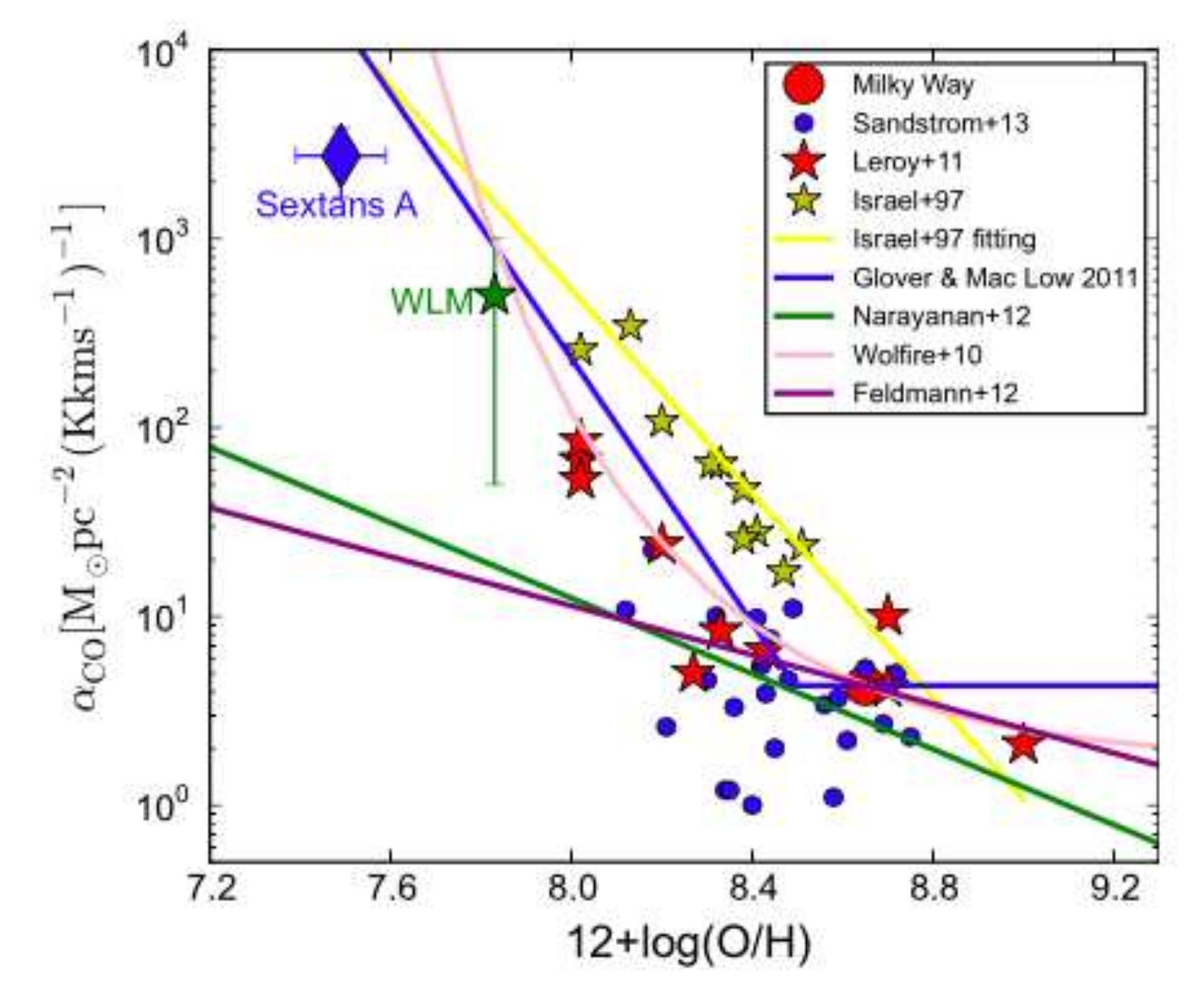}
  \label{fig:test2}
\end{minipage}
\caption{{\it Left:} Star formation rate surface density ($\Sigma_{SFR}$) vs. gas surface density ($\Sigma_{gas}$) updated from \cite{cormier14}. Dwarf galaxies are star symbols: H$_2$, determined from CO or HI. Starburts and spirals are the black squares from \cite{kennicutt98} . The diagonal line is the fit to the Schmidt-Kennicutt law with power index of 1.4.  If the H$_2$ for the dwarf galaxies is accounted for via the CO observation with a standard CO-to-H$_2$ connversion factor, they appear to have very efficient star formation. {\it Right:} CO-to-H$_{2}$ conversion factor ($\alpha_{CO}$) vs 12+log(O/H) (\cite[Shi \etal\ 2015]{shi15}). Symbols are observations from the literature. The lowest metallicity galaxies for which CO is detected are Sextans A and WLM.  The lines show various theoretical or empirical models. Lower metallicity conversion factors are uncertain.}
\label{co}
\end{figure*}

Dwarf galaxies (low-surface brightness, irregular, blue compact) are rather gas-rich, hosting large quantities of HI gas that often extend much further than the optical body. In the cold ISM, and especially the dense phases that fuel star formation, gas is expected to be mainly in molecular form (H$_2$). Cold H$_2$ is difficult to observe directly and its often-used tracer, carbon monoxide (CO), has very low abundance and therefore its emission is also difficult to observe in dwarf galaxies (e.g., \cite[Tacconi \etal\ 1987]{tacconi87}, \cite[Sage \etal\ 1992]{sage92}, \cite[Israel \etal\ 1995]{israel95}, \cite[Taylor \etal\ 1998]{taylor98}, \cite[Leroy \etal\ 2005]{leroy05}, \cite[Leroy \etal\ 2009]{leroy09}, \cite[Wong \etal\ 2011]{wong11}, \cite[Cormier \etal\ 2014]{cormier14}, \cite[Hunt \etal\ 2015]{hunt15}, \cite[Rubio \etal\ 2015]{rubio15}, \cite[Schruba \etal\ 2017]{schruba17}), although observing CO remains challenging at the very low metallicities, even with \textit{ALMA} (e.g., \cite[Cormier \etal\ 2017]{cormier17}). Theoretically, this is understood as CO photo-dissociation occuring on large spatial scales. However, H$_2$ can still be present in amounts that remain poorly quantified (Section \ref{codark}).The scarcity of detected CO, if this alone translates into the total H$_{2}$, would mean that the star-forming dwarf galaxies have a high star formation efficiency and are outliers on the Schmidt-Kennicutt relation (Figure \ref{co}). The consequence of the difficulty in detecting CO in dwarf galaxies is uncertainty in converting CO to H$_{2}$ (Figure \ref{co}). HI is plentiful, but CO is scarce.\\

\subsection{Uncovering the CO-dark gas}
\label{codark}
Calibration of the CO-to-H$_{2}$ conversion factor in low metallicity galaxies has been a perplexing issue (e.g., \cite[Schruba \etal\ 2012]{schruba12}, \cite[Bolatto \etal\ 2013]{bolatto13} and references within).  The combination of lower dust abundance and intense radiation field in star forming dwarf galaxies has important effects on the structure and physical properties of the surrounding PDR/molecular clouds. Molecular clouds can suffer deeper photodissociation leaving small CO cores and characteristically porous ISM conditions, further channelling the hard radiation field. Due to the self-shielding effect of H$_{2}$, an important molecular gas reservoir can exist outside the CO-emitting region (CO-dark gas; e.g., \cite[Wolfire \etal\ 2010]{wolfire10}). Such conditions can profoundly alter the structure of the HII region/PDR/molecular cloud.
 
\begin{figure}[t]
\begin{center}
\includegraphics[width=0.6\textwidth]{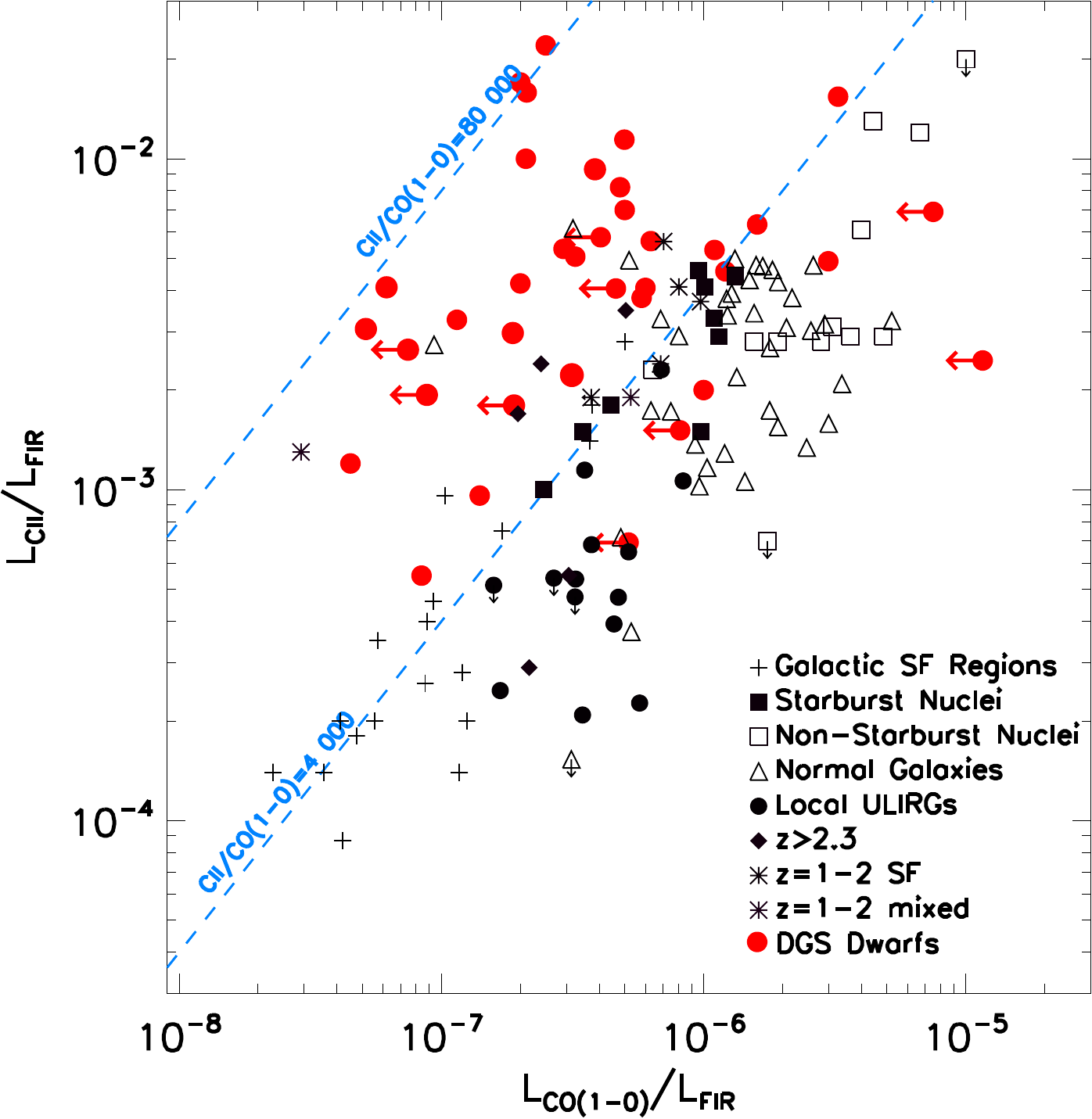} 
\hfill
 \caption{L$_{CII}$/L$_{FIR}$ vs L$_{CO}$/L$_{FIR}$ observed in a wide variety of objects. The data for the DGS dwarf galaxies are from \cite{cormier15}. The diagonal lines are constant L$_{CII}$/L$_{CO}$ (Madden \textit{et al.} in prep.). Note the elevated L$_{CII}$/L$_{CO}$ for the dwarf galaxies, much higher even than the metal-rich starburst galaxies.}
\label{ciico}
\end{center}
\end{figure}

 The presence of this important reservoir of molecular gas was first surmised via excessively bright [CII] observations in dwarf galaxies (\cite[Poglitsch \etal\ 1995]{poglitsch95}, \cite[Israel \etal\ 1996]{israel96}, \cite[Madden \etal\ 1997]{madden97}).  L$_{[CII]}$/L$_{CO}$, thought of as an indicator of star formation activity (\cite[Stacey \etal\ 1991]{stacey91}), is at least an order of magnitude higher in dwarf galaxies than metal-rich star-forming galaxies (Figure \ref{ciico}). In our galaxy, this CO-dark gas component has been traced via gamma ray observations (\cite[Grenier \etal\ 2005]{grenier05}) and dust observations (\cite[Planck collaboration \etal\ 2011]{planck11}) and subsequently via Herschel [CII] observations (e.g., \cite[Pineda \etal\ 2013]{pineda13}, \cite[Langer \etal\ 2014]{langer14}). 

The molecular gas reservoir has also been estimated by the IR dust emission. From the IR emission, the dust mass can be determined and with an estimated D/G the total gas mass can be estimated. The difference between the measured HI column density and the total gas mass will give the total H$_{2}$. Such approach has been used in Local Group Galaxies (e.g. \cite[Leroy \etal\ 2007]{leroy07}, \cite[Leroy \etal\ 2011]{leroy11}, \cite[Bolatto \etal\ 2011]{bolatto11}), and systematically high conversion factors are found for dwarf galaxies, as expected for their lower metallicity ISM (see \cite[Bolatto \etal\ 2013]{bolatto13}).

We demonstrate how the multi-phase modeling described above can be used to quantify the CO-dark reservoir in dwarf galaxies via the FIR [CII] observations.
The following applies to entire galaxies but there are also efforts to characterize the cold gas reservoirs on resolved scales, from [CII] and [CI] observations with \textit{ALMA, APEX}, and \textit{SOFIA} (e.g., \cite[Requena-Torres \etal\ 2016]{rt16}, \cite[Jameson \etal\ 2018]{jameson18}, \cite[Okada \etal\ 2018]{okada18}).
 
\subsection{Quantify the CO-dark gas with [CII] observations}
The multi-phase models of the dwarf galaxies (Section \ref{modeling}), which have been successfully confronted with the numerous observations of dwarf galaxies (\cite[Cormier \etal\ 2015]{cormier15}, \cite[Cormier \etal\ 2018]{cormier18}), also provide the total H$_{2}$ as an outcome of the modeling for each of the galaxies. The difference between the model estimate of the total H$_{2}$ and the H$_{2}$ determined from CO observations, using a standard conversion factor, quantifies the CO-dark gas reservoir. It is essentially the H$_{2}$ that lies outside the CO-emitting region, where the CO has been photo-dissociated, while the H$_{2}$ survives via self-shielding.

\begin{figure}[t]
\begin{center}
\includegraphics[clip,trim=5mm 0 5mm 0,width=2.62in]
{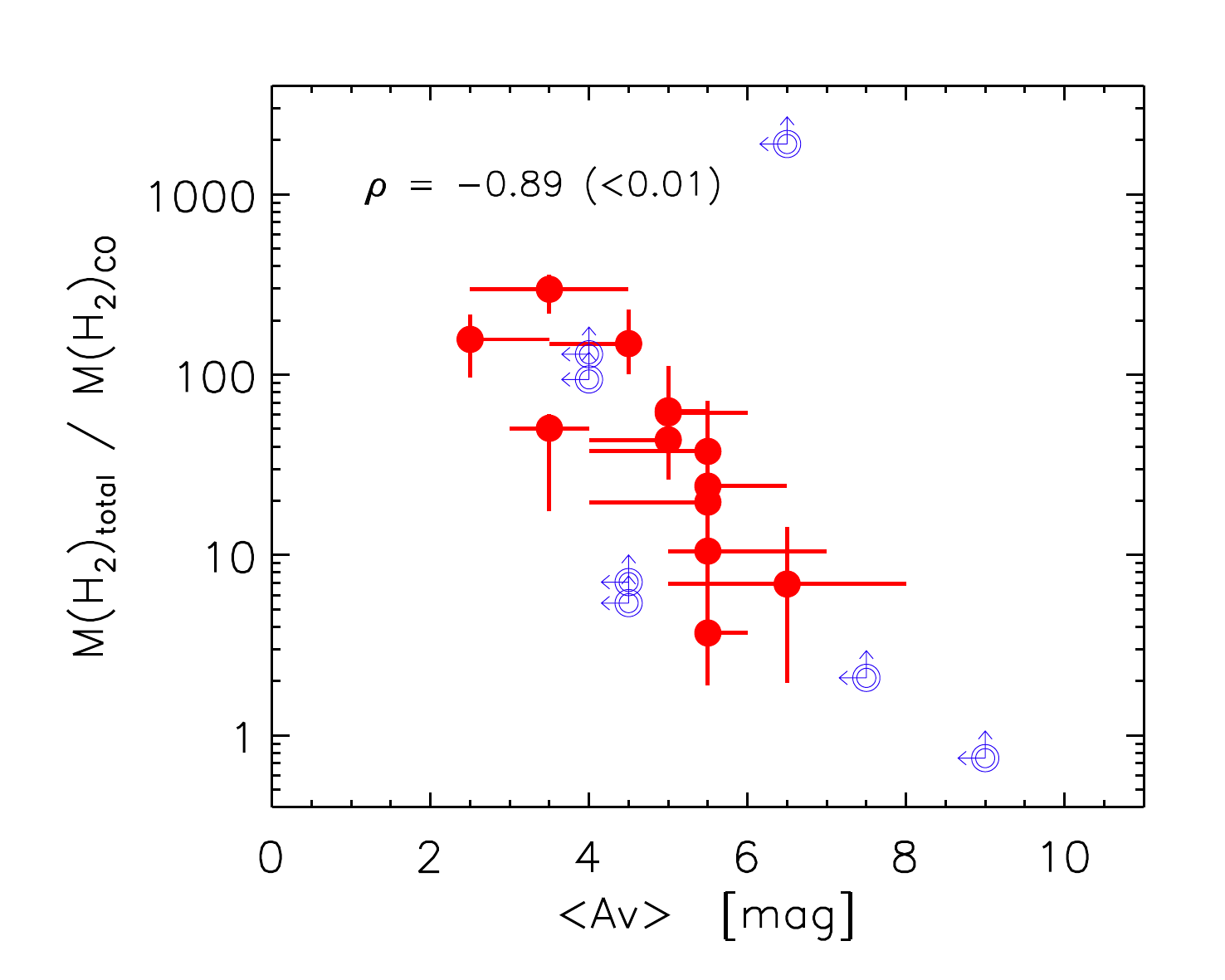} \hfill
\includegraphics[clip,trim=5mm 0 5mm 0,width=2.62in]
{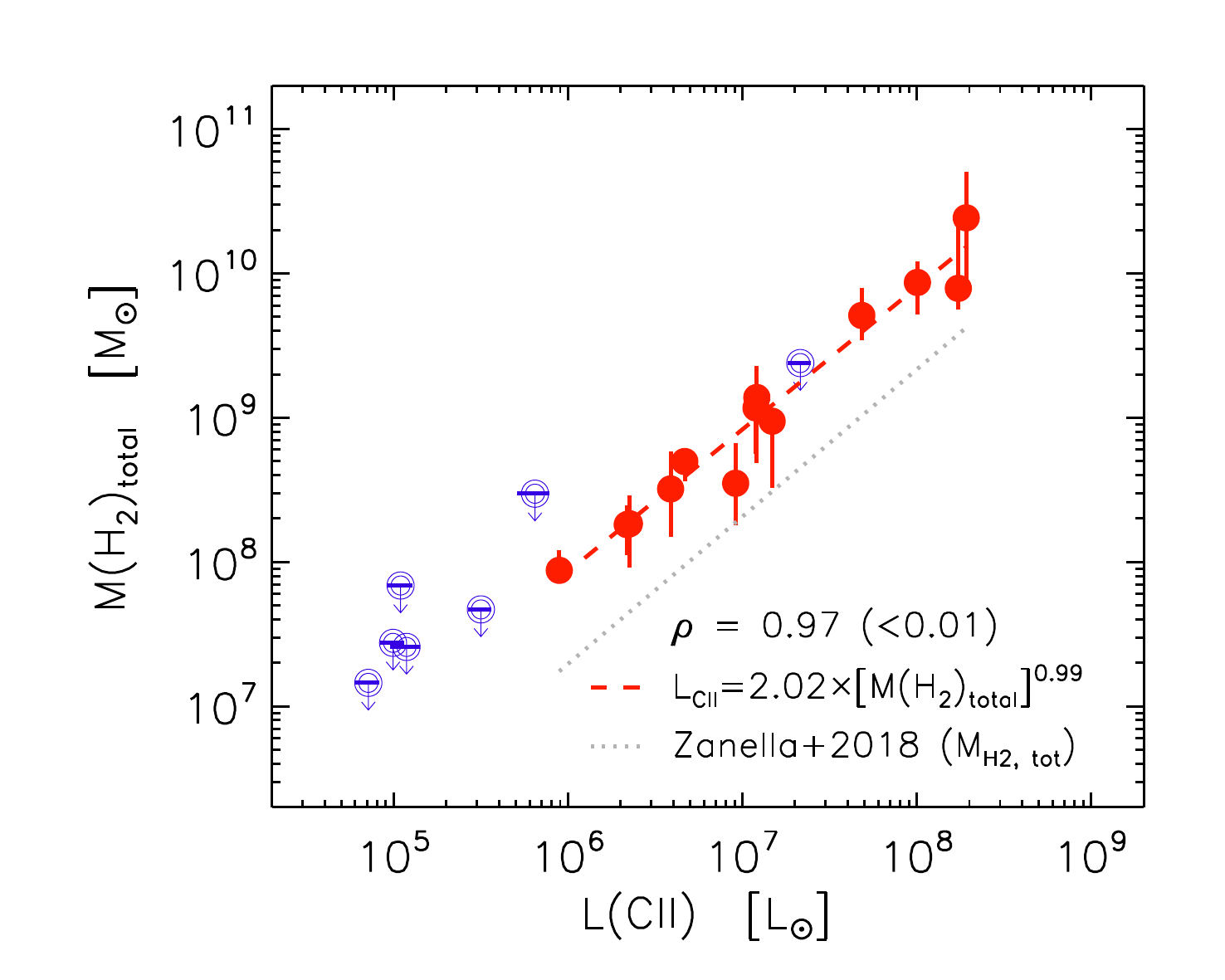}
 \caption{{\it Left:} The CO-dark gas mass fraction (ratio of total mass of H$_2$ over the mass of H$_2$ traced by CO) is a strong function of the mean $A_{\rm V}$ of the neutral ISM. {\it Right:} Correlation between the total molecular gas mass (CO-bright and CO-dark) and the [CII] luminosity for the DGS galaxies with [CII] and CO observations (Madden et al. in prep.).}
\label{fig:obscodark}
\end{center}
\end{figure}

For the Dwarf Galaxy Survey, the CO-dark gas reservoir dominates the molecular gas reservoir and clearly, CO misses most of the H$_{2}$. While there is no correlation of CO-dark gas mass fraction with metallicity, the global value of $A_{\rm V}$ regulates the fraction of CO-dark gas (Figure~\ref{fig:obscodark}); the $A_{\rm V}$ impacts the transition between C$^{+}$ and CO. This model reproduces the extreme observed L$_{CII}$/L$_{CO} $ (Figure \ref{ciico}).  As a consequence, the  L$_{CII}$/L$_{CO}$ ratio is an excellent tracer of the total H$_{2}$ (Madden et al. in prep.) and we find the following relation:
\begin{equation}
    {\rm 
    L_{[CII]}/L_{CO(1-0)} = -3.5 \times
    (M(H_2)_{total} / M(H_2)_{CO})^{1.2}
    }
\end{equation}

\begin{figure}[t]
\begin{center}
\includegraphics[clip,trim=1mm 0 5mm 0,width=2.62in]
{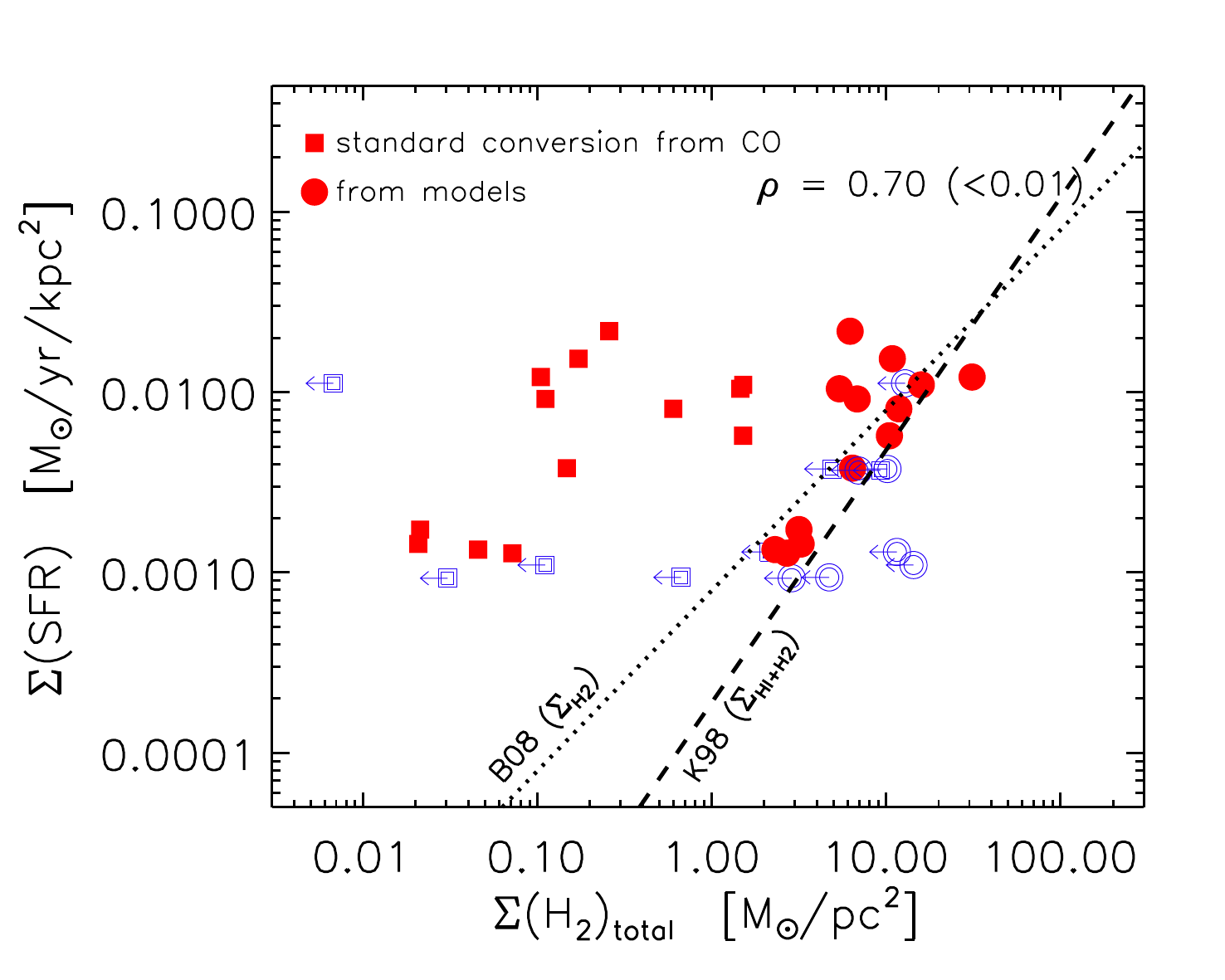} \hfill
\includegraphics[clip,trim=2mm 0 5mm 0,width=2.62in]
{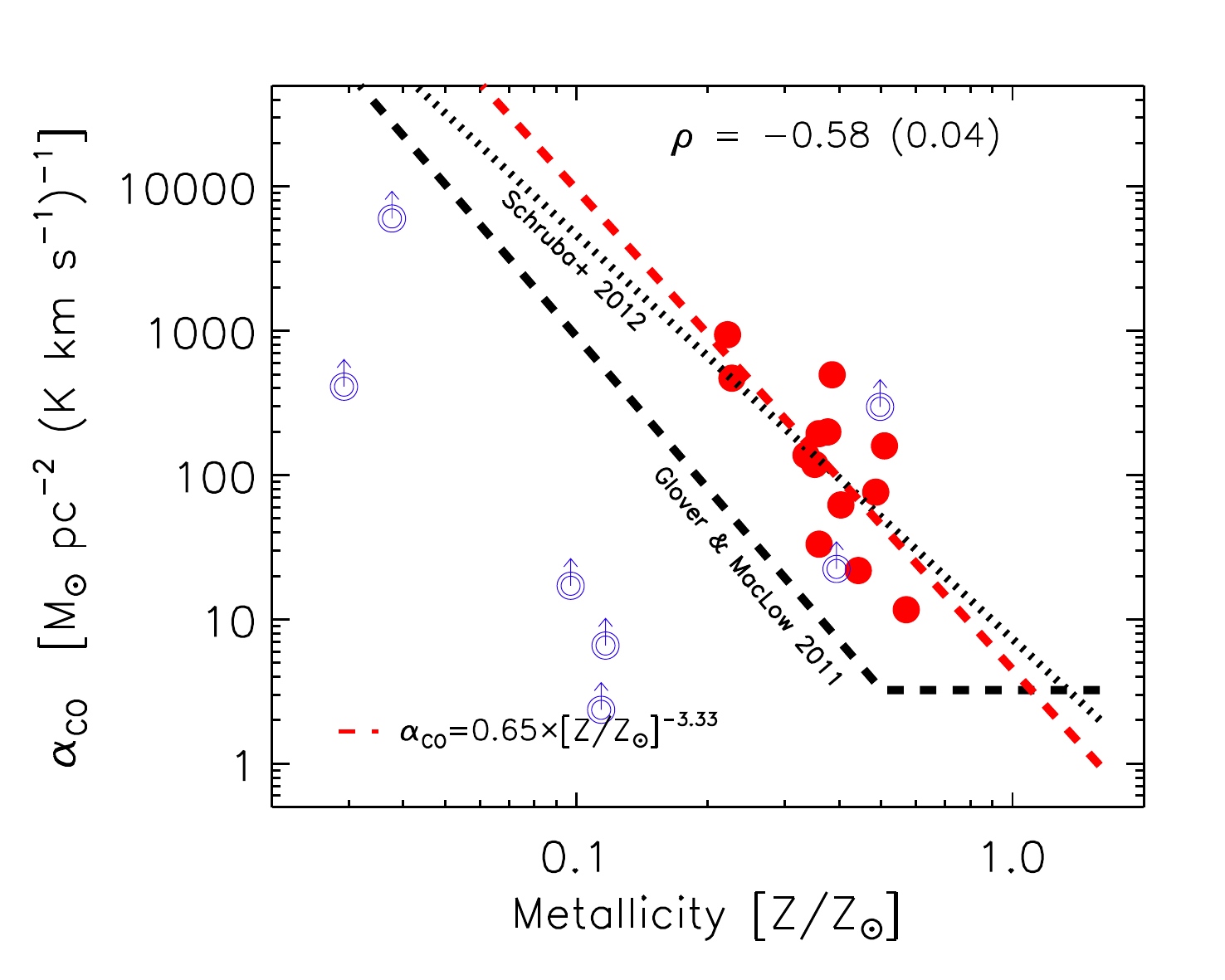}
 \caption{{\it Left:} Revisited Schmidt-Kennicutt relation for the dwarf galaxies of the DGS. Using the total H$_2$ gas masses from the models (i.e. including both CO-bright and CO-dark gas reservoirs), the dwarf galaxies fall closer to the main relation for normal galaxies. {\it Right:} Correlation of the conversion factor $X_{\rm CO}$ with metallicity. We find a very steep relation for the DGS galaxies (Madden et al. in prep.).}
\label{fig:obsxco}
\end{center}
\end{figure}

When we then revisit the Schmidt-Kennicutt relationship, we now see that taking into account the {\it total} H$_{2}$ for the dwarf galaxies, the galaxies shift toward the expected relationship (with some scatter; Figure~\ref{fig:obsxco}). The conversion of CO-to-H$_{2}$ can then be reformulated, to take into account the CO-dark gas:
\begin{equation}
    \alpha_{\rm CO} = 0.65 \times (Z/Z_{\odot})^{-3.3}~~
    [{\rm M_{\odot}\,pc^{-2}\,(K\,\,km\,s^{-1})^{-1}}]
\end{equation}
This new relation found for the DGS galaxies is a very steep function of metallicity (Figure~\ref{fig:obsxco}), steeper than previously estimated (from other methods).

\section{Conclusion and perspectives}
\label{conclusion}

Star-forming dwarf galaxies are unique laboratories to study the effects of star formation on the observed gas and dust properties. The lowest metallicity dwarf galaxies have a scarcity of dust, and require grains to grow in the ISM until 12+log (O/H) $\sim$ 8, when they can efficiently turn metals into dust. The low dust abundance and hard and intense radiation fields impact the ISM on galaxy-wide scales. This can be see via the MIR and FIR fine structure emission lines along with multi-phase modelling. Ionized gas tends to fill a large volume of the ISM of dwarf galaxies, while the neutral gas has a low covering factor. This can explain the dearth of CO leading to uncertainty in quantifying the H$_2$ reservoir using the CO line emission. The L$_{CII}$/L$_{CO}$ ratio is an excellent tracer of the total H$_{2} $ and can be a particularly useful calibrator for molecular gas for high redshift galaxies.

As the community is acquiring copious data sets (multi-wavelengths and multi-scales) of low metallicity galaxies and pushing these observations to higher red-shifts, multi-phase modelling of the gas and dust is an appropriate means to get to the structure and physical properties of the different phases within galaxies. In the future, it will be important to build on these studies and add complexity to the models, in particular, by integrating dynamical effects and realistic geometries.

Greater sensitivity in the MIR and FIR to obtain the required variety of diagnostics of the different phases of the lowest metallicity galaxies requires a cold telescope. This could be achieved with the \textit{SPICA} telescope, a possible M5 ESA-JAXA mission, which has just started Phase A, and will hopefully be the final candidate selected in 2021, bringing us unprecedented sensitivity at MIR and FIR wavelengths. In wavelength space, \textit{SPICA} falls between the \textit{JWST} and \textit{ALMA}, which provide the spatial resolution to zoom into the details of the small scale structures. All of these different views of the ISM of dwarf galaxies are necessary to piece together a coherent story on the evolution of the dust and gas in dwarf galaxies and to eventually understand their role in the cosmic history.


%

\end{document}